\documentclass[10pt,onecolumn]{IEEEtran}
\usepackage{graphicx}
\usepackage{url}
\usepackage{cite}
\usepackage{authblk}
\usepackage{booktabs}
\usepackage{subfigure}
\usepackage{algorithm}
\usepackage{mathtools}
\usepackage{color}
\begin{document}

\title{Caveat emptor: the risks of using big data for human development}

\author[1,2]{Siddique Latif}
\author[1]{Adnan Qayyum}
\author[1]{Muhammad Usama}
\author[1]{Junaid Qadir}
\author[3]{Andrej Zwitter}
\author[4]{Muhammad Shahzad}

\affil[1]{Information Technology University (ITU)-Punjab, Pakistan}
\affil[2]{University of Southern Queensland, Australia}
\affil[3]{University of Groningen, Netherlands}
\affil[4]{National University of Sciences and Technology (NUST), Pakistan}

\maketitle

\begin{abstract}
\sloppy
Big data revolution promises to be instrumental in facilitating sustainable development in many sectors of life such as education, health, agriculture, and in combating humanitarian crises and violent conflicts. However, lurking beneath the immense promises of big data are some significant risks such as (1) the potential use of big data for unethical ends; (2) its ability to mislead through reliance on unrepresentative and biased data; and (3) the various privacy and security challenges associated with data (including the danger of an adversary tampering with the data to harm people). These risks can have severe consequences and a better understanding of these risks is the first step towards mitigation of these risks. In this paper, we highlight the potential dangers associated with using big data, particularly for human development. 
\end{abstract}


\section{Introduction}

Over the last decades, widespread adoption of digital applications has moved all aspects of human lives into the digital sphere. The commoditization of the data collection process due to increased digitization has resulted in the ``data deluge'' that continues to intensify with a number of Internet companies dealing with petabytes of data on a daily basis. The term ``big data'' has been coined to refer to our emerging ability to collect, process, and analyze the massive amount of data being generated from multiple sources in order to obtain previously inaccessible insights. Big data hold the potential and enables decision makers to ascertain valued insights for enhancing social systems, tracking development progress, understanding various aspects of policies, adjusting program requirements, and making decisions based on evidence rather than intuitions. 

In particular, the recent advances in ML and AI techniques have revolutionized intelligent data analytics which has resulted in tremendous interest in the adoption of big data-driven technologies for sustainable human development. Many organizations and government institutions are increasingly experimenting such solutions in different fields, e.g., healthcare, education, intelligence, fraud, and crime prevention just to name a few. There are tremendous scientific potentials and possibilities with big data \cite{ali2016big}: however, big data is not a silver bullet and we can only ignore at our own peril the hard-earned statistical lessons on measurement bias, data quality, and inference variation that have been earned through a hard toil. While most writing on data is enthusiastic, new work has started emerging that has begun to show how big data can mislead and be used detrimentally \cite{o2016weapons,boyd2012critical, helbing2017will}. More than 50 years of research into artificial intelligence and statistical learning has shown that there is no free lunch---i.e., there are no universally applicable solutions \cite{wolpert1996lack} and that there are always trade-offs involved. 

The use of \textit{big data for development} (BD4D) is transforming society, from diagnosis of illness to education. It has improved the efficiency and effectiveness of disaster management systems by utilizing real-time community information \cite{qadir2016crisis}. In these applications, data is emerging as a new economic resource and used by companies, governments, and even individuals to optimize everything possible \cite{mcafee2012big}. But the dark side of big data is that it erodes privacy and threatens freedom when it is used for human development. The data can be of poor quality,  biased, misleadingly, and even more damning that fail to capture purports. Similarly, when big-data predictions about individuals are used to punish people on their propensities but not their actions. This denies independence and frets human dignity \cite{john2014big} and causes greater sectarianism where people will be testified with more data, more proof, more conviction but a lesser degree of tolerance for dissenting views   \cite{silver2012signal}.

\footnote{© 2019 IEEE.  Personal use of this material is permitted.  Permission from IEEE must be obtained for all other uses, in any current or future media, including reprinting/republishing this material for advertising or promotional purposes, creating new collective works, for resale or redistribution to servers or lists, or reuse of any copyrighted component of this work in other works.}



Despite the great excitement around the big data trend, big data is also pegged with a lot of criticism from interdisciplinary works due to some common problems. These issues become more serious and need more attention when big data is used for human development. Various studies have articulated common issues that concern and affect interdisciplinary research using big data analytics \cite{ekbia2015big,sivarajah2017critical,mittelstadt2016ethics,hashem2015rise}. However, critical analysis of BD4D is missing in the literature. Therefore, in this paper, we discuss the caveats of big data analytics when it is used for human development purposes. The purpose of this article is to present a critical synthesis of the diverse literature and put together various underlying issues that affect BD4D significantly. In addition to discussing the critiques of big data, this paper also discusses various potential remedies that can help in the mitigation of these big problems of big data not only for BD4D. 

\section{General Issues in Using Data for Decisions}

Today, big corporations are investing their resources in utilizing big data technology to uncover all hidden patterns, important correlations, customer preferences, market trends, and other useful information to boost their production efficiency by providing data-driven products and services. Despite these great opportunities, big data has come with a raft of potential challenges and pitfalls that need to be considered. Some of them are discussed next. 

\subsection{Big data is Reductionist}

Big data provides a granular view of extremely complex and historically unprecedented problems by using particular vantage points from the available data. It only focuses on specific types of questions, and seek their answers in a reductionist way by seemingly ignoring concrete contextual realities about human societies and places. It is also crude and reductionist in its techniques and algorithms by sacrificing complexity, specificity, and deep contextual knowledge \cite{kitchin2014big}. In this way, analyses are reduced to simple outputs that critically ignore the touting potentials of big data. Reductionism is extremely helpful in exploring various mechanisms from real-life, but the translation of these outputs into real implementations is very difficult. 

\subsection{Big Data is Not Neutral Nor Is It Objective}


One of the compelling aspects of big data is the assumption that it is eliminating human subjectivity and bias. Although the idea of ``baggage-free'' learning is seductive, data alone is not enough since there is no learning without knowledge as a practical consequence of the ``no free lunch'' theorem. Big data analytics is what mathematicians call an ill-posed problem thus no unique model exists\footnote{A famous statistical quote by George Box is, ``All models are wrong, some are useful.''}; the only way to resolve some ill-posed problems is to make additional assumptions. These assumptions are subjective or \textit{``opinions embedded in Math''} as described by Cathy O'Neil in her best-selling book \cite{o2016weapons}. Therefore while data is often considered neutral, however data collection techniques and analysis methods are designed by humans based on their knowledge, experience, and beliefs. 

The German philosopher Heidegger wrote, ``\textit{Everywhere we remain unfree and chained to technology, whether we passionately affirm or deny it. But we are delivered over to it in the worst possible way when we regard it as something neutral}.'' Big data scientists make computations using data and present results about people in terms of numbers or mathematical relations based on layers upon layers of assumptions. Data scientist use this data and try to organize it into objects, relations, and events, and in doing so invariably analyze the data in the backdrop of their subjectivity and partiality. Moreover, it is challenging to ensure fairness in ML algorithms due to significant statistical limitations and it is possible that enforcing fairness can harm the very groups that these algorithms were supposed to protect \cite{corbett2018measure}. 

\subsection{Opaque Black-box Models}

Many of the modern mathematical models used for big data analysis are opaque, and unregulated \cite{o2016weapons}. These models are often used inappropriately and can scale up the biases thereby producing misleading results. Such models have even been referred to as ``weapons of maths destruction'' (WMD) due to their inscrutable black box nature. These WMDs can cause havoc/harm and tend to punish the poor due to vicious feedback loops \cite{o2016weapons}. The reliance on opaque algorithmic black boxes for decision making is even more problematic when we consider how easy it is for adversaries to attack ML models. New research attacks have emerged in which adversaries can trick ML models to make decisions as they desire---something that has huge consequences for a society in which human decisions are automated and made by machines.






 
\subsection{Big Data's Big Bias Problem}

Most big datasets contain hidden biases both in collection and analysis stage. These biases in data can create an illusion of reality. These biases are hard to undo and their elimination have unintended consequences on the results. Four major biases of big data are described as follows. 

\vspace{2mm}
\subsubsection{Sampling Bias} 

When the samples are partial, selective, and not random, the patterns of omitted information may influence the structures discovered in the data. Such samples will be unable to accurately predict the outcomes \cite{price2015selection}. In this way, analysis using big data will present a bounded scope with very lesser means to prove its validity. Examples of studies that obviously contain such bias are sentiment analysis on social media, safety monitoring using social media, population statistics, and tourism statistics.

\vspace{2mm}
\subsubsection{Activity Bias} 

This is another bias that is usually encountered in web data \cite{lewis2011here}. This arises from the time-based correlation of users' activities across diverse websites. Because most of the users visit a website for a while and never return during the measurement period. This can be explained using the example of the model of Ginsberg et al. \cite{ginsberg2009detecting} for predicting flu cases in the US using 50 million Google search terms suffered from activity bias where the model has started over-predicting the flu in the US \cite{lazer2014parable,butler2013google}.

\vspace{2mm}
\subsubsection{Information Bias} 

It refers to the delusion that more information always results in better decisions. Data scientists give too much significance on the volume of data and ignore other alternative opinions and solutions. The irony of this belief is that fewer data can give better decisions in some situations. In various cases, a simple rule of thumb can work better than complex analysis. There are situations where ignorance from very large data and required calculation can provide more accurate and faster results because large biased data just magnify the errors. Having more information is not always more desirable.


 
\vspace{2mm}
\subsubsection{The Inductive Bias (Assuming Future Will be Like Past)}

In big data analysis, a common erroneous belief is that the future has a direct correlation with the past. This can undermine a good decision. When we look backward for the longer view, the far-distant past will be shriveled into the meaningless importance and becomes irrelevant for systems to predict future patterns \cite{taleb2005fooled}. In many cases, such analyses and inferences are not only invalid but also misleading and unhelpful.

\subsection{Fooled by the Big Noise}


Nate Silver has reported in his book \cite{silver2012signal} that data is increasing globally by 2.5 quintillion bytes each day, but useful information is of course not increasing commensurately. This means that most of the data is just noise and that noise is increasing faster than the useful information. A related problem is that of the curse of dimensionality due to which the more dimensions one works with, the less effective standard computational and statistical techniques become, which have serious repercussions when we are dealing with big data. As data get larger, the complexity, deviations, variance (or noise) and the number of potential false findings grow exponentially compared to the information (or signal). This spurious rise in data can mislead us to fake statistical relationships.

Another point of concern is that a plethora of hypotheses are tested using a single data in big data analytics, which opens the door for \textit{spurious correlations}. When a large number of hypotheses are tested, it becomes highly likely that some false results will become statistically significant, which can mislead big data practitioners if they are not careful. The proper way to formulate a hypothesis is before the experiment not after it. It is necessary to understand that the statistical significance under an inaccurate method is totally specious---i.e., significance tests do not protect against ``data dredging'' \cite{smith2002data}.

Apart from the problem of spurious correlations, big data analysts can also become guilty of \textit{cherry picking}. In this phenomenon, scientists tend to focus on finding support for their hypotheses and beliefs and overlooking other evidence. As a result, they only present positive results of their experiments that endorse their hypothesis or argument instead of reporting all of the findings \cite{morse2010cherry}. In this way, big data provide a relatively small advantage to improve public strategies, policies, and operations with minimum public benefit \cite{brooks2013philosophy}.

\subsection{Are Big Data Predictions Generalizable?}

In big data analytics, datasets are mostly gathered from observational sources instead of a strict statistical experiment. This poses the question of the generalizability of the insights learned from this data notwithstanding the large size of the dataset. Even though we know that statistically speaking the discrepancy in an estimate decreases with an increase in the sample size as Bernoulli proved, but the messiness of real-world data in terms of incompleteness, imbalance, hidden bias, rareness, large variances, outliers, and non-i.i.d nature\footnote{The i.i.d. assumption assumes that the data is statistically independent and identically distributed}, means that simply getting more data is not sufficient. Disregard for the messiness of the real-world data can mislead us into problems such as multiple testing, regression to the mean, selection bias, and over-interpretation of causal associations \cite{o2016weapons}. Also we must note that certain things are not precisely predictable (e.g., chaotic processes, complex systems, the so-called black swans) no matter how much data is available. 

In BD4D applications, large administrative databases may have obscure/indeterminate data quality, limited information about confounding variables, and sub-optimal documentation of the outcome measures, and careful attention should be given to the generalizability of the learned insights.

\section{Data in the Field of Development}


\subsection{Ethical Use of Data and Privacy Issues}

Big data opens frightening opportunities for unscrupulous people, group, government, or organizations to use users' personal data against them for evil purposes as various facets of human lives are being digitized \cite{boyd2012critical}. Privacy depends on the nature and characteristic of data, the context in which it was created, and the norms and expectations of people \cite{boyd2012critical}. Therefore, it is necessary to situate and contextualize data in such as way that minimizes privacy breaches without affecting its usefulness. 

Big data can bring more transparency both for individuals and organizations by introducing user anonymization but there are numerous examples when data is thought to be anonymous but its combination with other variables resulted in unexpected reidentification \cite{kloumann2014community,shelton2015social,acquisti2011privacy}. For example, it has been proved that four spatiotemporal points are sufficient to uniquely identify 95\% of the individuals in a dataset where the location of an individual is specified hourly with a spatial resolution given by the carrier's antennas \cite{de2013unique}. In another similar study, Latanya Sweeney \cite{sweeney2000simple} argued that 87\% of all Americans could be uniquely recognized using only three bits of information: zip code, birth date, and sex. The datasets being released now days are anonymized by applying ad-hoc de-identification methods. Therefore, the possibility of reidentification depends heavily on the advancement of reidentification methods and availability of the auxiliary datasets to an adversary. Thus, the chances of privacy breaches in the future are essentially uncertain.

The all-important consideration that should be accounted for is that how the collection of data may affect a person's well-being, dignity, and harmony. It is important to ensure the ethical use of data explicitly when it is used for human development. We can actively minimize the risk of inflicting harm by 1) revealing confidential information; 2) malicious use of data (potentially by unauthorized agents); or 3) altering their environment.  

\subsection{Quality of statistics---numbers are soft, and incentives matter}


Data is not simply information that is harvested from an objective context. It should be institutionally, culturally, and socially shaped product. Collecting a good quality data for development is in practice very costly as we need resources and manpower for data collection, maintenance, and monitoring \cite{jerven2017much}. For instance, the aggregate statistic GDP in many sub-Saharan African countries is often measured approximately since most African countries are simply unable to collect all the information needed to calculate GDP accurately and changes are approximately inferred from rainfall figures or population growth \cite{jerven2013poor}. The use of this data can lead to distorted or misleading policy decisions by development agencies and governments and can contribute to the underestimation of Gross Domestic Product (GDP) and bewildering fallouts such as the following:

\begin{enumerate}

\item In November 2010, Ghana's Statistics Service announced estimated GDP that was off from its true value resulting in an upward adjacent of a whooping 60\% which was enough to change Ghana's status from a low-income country to a lower middle-income country; 

\item Similar adjustments were made in Nigeria in April 2014 where the rise was almost 90\%, which causes the total GDP of sub-Saharan Africa to rise by 20\% \cite{jerven2013poor}. 
\end{enumerate}

But where did these overnight growths come from? And what to say about the fate of the analysis and policies of the various policymakers and development professionals that were based on the previous miscalculated data---and the cost of this misplaced analysis? As the official statistics are often missing, incomplete, dated, or unreliable. Given such concerns, researchers have argued that the numbers cannot be taken at face value for the development data from a place like Africa resulting in what has been termed as the ``\textit{Africa's statistical tragedy}'' \cite{jerven2013poor}. 


\subsection{The Pitfalls of Self-Monitoring of States} 


In the modern world, almost every government tightly control what kind of information is disseminated about the state. An important aspect of the self-monitoring of states is that it often becomes a farce because people cannot keep their own score. The official statistics measured by the government must be accepted with a pinch of salt since governments are spinning these numbers in ways that project the country's progress positively since the legitimacy of the state depends on the popular understanding of the country's progress. The anthropologist James Scott in his book ``Seeing Like a State'' \cite{scott1998seeing} describes the ways in which governments, in their fetish for quantification and data, end up making people’s lives miserable rather than better. As development projects initiated by the states are aimed at improving human lives, an appropriate measure for the development must be how much effort is expended on minimizing the risks associated with data collection and analysis.

\subsection{Human Systems Have Complex Loops With Predictions Being Self-Fulfilling}


The human social systems are complex adaptive systems with multi-loop nonlinear feedback in which actions performed for some purpose can lead to quite different unexpected and unintended consequences. We will use two examples to illustrate the (1) unintended consequences and (2) self-fulfilling nature of interventions in complex social systems. 

For our \textit{first} example, consider the \textit{Cobra effect} as an example of an incident in which an intervention crafted to ameliorate a problem actually aggravated it by producing some unintended consequences. During the British colonial rule of India, the government devised a bounty system for combating the rise of venomous cobras. The system worked successfully initially and lots of snakes were killed for the reward but entrepreneurs soon figured out that they could make money by farming cobras and killing more of them. The government on learning about this scrapped the system but ended up with a situation in which there were more cobras after the intervention than before. 

For our \textit{second} example, consider the \textit{Paper Town} effect. In the 20th century, a famous map of New York (NY) was created by cartographers Lindberg and Alpers, who cleverly embedded a fake city Agloe into their map. Agloe NY was not a real town. It was a paper town—a booby trap to catch plagiarizers. A few years after Lindberg and Alpers set their map trap, the fake town appeared on another map by the cartographer Rand McNally map, prompting the two mapmakers to sue for copyright infringement. Eventually it was discovered that a real town called Agloe had in fact emerged in New York\footnote{\url{https://en.wikipedia.org/wiki/Agloe,_New_York}} since users of the Linberg Alpers map thought that Agloe that once existed must have gone missing and rebuilt it and that Rand McNally may not have after all have plagiarized Lindberg and Alpers.

\subsection{Missing Data Problem}


Missing data is a big problem for development statistics. It has been reported by Jerven that around half of the 82 low-income countries have one or partial poverty survey within the past decade \cite{jerven2013poor}. Kaiser Fung, in his book \textit{Numbersense} \cite{fung2013numbersense}, discourages the assumption that we have everything and says: ``\textit{N = All is often an assumption rather than a fact about the data}''. Due to missing data problem, any ranking of such countries based on GDP only will be misleading because of the uneven use of methods and access to data. Handling missing data is very important for data mining process as missing observations can significantly affect the performance of the model \cite{mirkes2016handling}. Therefore, analysts should handle the missing patterns by employing appropriate methods to cope with it and to avoid the ``\textit{streetlight effect}''. The streetlight effect is the trend adopted by researchers to study what is easy to study. The streetlight effect is a major issue which restricts big data findings to be realistically useful for human developments---especially when findings are yielded using user-generated and easily available data.

\section{Some Remedies}



\subsection{Interpretable AI and Big Data Analysis}


With the wide adoption of deep learning and ensemble methods, modern AI systems have become complex and opaque and increasingly operate as a black box. Although these black-box models are producing outstanding results it is very difficult to trust their predictions due to there opaque and hard-to-interpret nature. This has been dubbed by some as the \textit{AI's interpretability problem}. We can define interpretability as the ability to describe the internal processes of a system (i.e. complex AI \& ML techniques) in such a way that they are understandable to humans \cite{gilpin2018explaining}. 

Interpretable AI can help ensure algorithmic fairness, bias identification, robustness, and generalization of big data based AI models. Interpretation of AI-based decision-making algorithms is also necessary to ensure smooth deployment of real-world intelligent systems. But the development of interpretable AI requires that the following questions be answered: 

\begin{enumerate}
    \item \textit{How to ensure accountability of model?} 
    \item \textit{How to ensure the transparency of the model output?}
    \item \textit{How to ensure the fairness of the model predictions?}
\end{enumerate}


Since BD4D is directly related to human development the big data based AI model used in these cases (i.e. healthcare system, judicial system etc.) must ensure high accuracy and interpretability. A possible remedy is to insist on using interpretable ML for high-stakes BD4D decisions and to utilize explanation methods for justifying the decisions, where explanation means the provision of visual or textual evidence of a certain features relation to AI models decision. As an example work in this space, Bach et al. \cite{bach2015pixel} proposed layer-wise relevance propagation method (LRP) which provides a visual contribution of each input feature in decision making. In another work, Ribeiro et al. \cite{ribeiro2016should} proposed two explanation methods namely \textit{locally interpretable model-agnostics} (LIME) and \textit{submodular pick locally interpretable model-agnostics} (SP-LIME). Interpretable AI for BD4D is still an open research avenue to explore for big data and AI community.

\subsection{Better Generalization}

In big data research, researchers make inferences by training models on a larger subset of data with the goal to fit the learned hypothesis on unseen data. In big data research, generalization is the procedure of spanning the characteristics of a group or class to the entire group or population. It allows us to infer attributes of an entire population without getting to know every single element in that population individually. The problem comes along when we wrongly generalize; or more precisely when we overdo it. The generalization fallacy occurs when statistical inferences about a particular population are asserted to a group of people for which the original population is not a representative sample. In other words, models overfit when it learned signal as well as noise from training data and unable to predict on unseen data. In order to avoid excessive generalization error, researchers should also check the scope of the results instead of extending scientific findings to the whole population. Regularization is helpful to avoid overfitting by reducing the number of parameters to fit model in high-dimensional data \cite{national2017refining}. It also prevents model parameters to change easily, which helps in keeping the focus of the model on the persistent structure. Apart from regularization, other techniques such as cross-validation, early stopping, weight sharing, weight restriction, sparsity constraints, etc. can also be used for reducing the generalization error based on the algorithm being used. 



\subsection{Avoiding Bias}

Big data tends to have high dimensionality and may be conflicting, subjective, redundant, and biased. The awareness of potential biases can improve the quality of decisions at the level of individuals, organizations, and communities \cite{kahneman2011before}. In a study of 1000 major business investments conducted by McKinsey, it was  found that when organizations worked to minimize the biases in their decision-making, they achieved up to 7\% higher returns \cite{lovallo2010case}. The biases associated with multiple comparisons can be deliberately avoided using techniques such as the Bonferroni correction \cite{weisstein2004bonferroni}, the Sidak correction, and the Holm-Bonferroni correction \cite{abdi2010holm}. Another source of bias is called \textit{data snooping} or \textit{data dredging}, which occurs when a portion of data is used more than once for model selection or inference. In technical evaluations of results, it is conceivable to repeat experiments using the same dataset to get satisfactory results \cite{white2000reality}. Data dredging can be avoided by conducting randomized out-of-sample experiments during hypotheses building. For example, an analyst gathers a dataset and arbitrarily segments it into two subsets, A and B. Initially, only one subset---say, subset A---is analyzed for constructing hypotheses. Once a hypothesis is formulated, it should then be tested on subset B. If subset B also supports such a hypothesis then it might be trusted as valid. Similarly, we should use such models that can consider the degree of data snooping for obtaining genuinely good results. 


\subsection{Finding Causality Rather than Correlations}

In most data analysis performed in the big data era, the focus is on determining correlations rather than on understanding causality \cite{Pearl2018}. For BD4D problems, we're more interested in determining causes rather than correlates and therefore we must place a premium on performing causal BD4D analysis since causally-driven analysis can improve BD4D decisions. Discovering causal relations is difficult and involves substantial effort, and requires going beyond mere statistical analysis as pointed out by Freedman \cite{freedman1991statistical} who has highlighted that for data analytics to be practically useful, it should be the problem-driven or theory-driven, not simply data-driven as Freedman says, using big data for development requires ``the expenditure of shoe leather'' to situate the work in the proper context. The focus on correlation has arisen because of the lack of a suitable mathematical framework for studying the slippery problem of causality until the recent fundamental progress made by Pearl \cite{Pearl2018}, whose work has now provided a suitable notation and algebra for performing a causal analysis. 






\subsection{Stress High-Quality Data Analytics Rather Than Big Data Analytics}


A better and thoughtful understanding of risks or pitfalls of big data is crucial to decrease its associated potential harms to individuals and society. There needs to be a stress on utilizing big data along with data collected through traditional sources can provide a deeper, clearer understanding of the problem instead of being fixated on only generating and analyzing large volumes of data. Although it is generally preferred to have more data, it is not always desirable, especially in the cases where data is biased. Another disadvantage of the large dataset is the cost in terms of processing, storage, and maintenance. However, some simple methods like \textit{sampling and/or resampling} enables us to extract the most relevant data from a larger chunk of data. Another very important aspect is to collect the desired data to properly design experiment rather than collecting all possible information. Specifically, in the field of human development and humanitarian action, there is no way around corroborating findings based on big data with intelligence gathered on the field level. This requires that international organizations and development actors actively increase the capacity for data collection and analysis also referred to as ``\textit{Humanitarian Intelligence}'' \cite{zwitter2016humanitarian}.






\subsection{User-Friendly \& Responsible Data Analytics}

Big data algorithms are not as trustworthy as we think since they draw upon data collected from a prejudiced and biased world.  Cathy O'Neil \cite{o2016weapons} describes how algorithms often perpetuate or worsen inequality and injustice and suggests that there should be laws and industry standards to ensure transparency for big data gathering and utilization. In particular, false, outdated, and taken out of context information may cause harm to an individual's autonomy and the use of such information should, therefore, be restricted. As a remedy to this issue, ``\textit{the right to be forgotten}'' enables data subjects to reassert control over their personal information. There might be fair audits of algorithms but first, the awareness of this issue to programmers is required as they share a disproportionate amount of responsibility in the design of big data-based algorithms. Since AI and ML algorithms are embedded in many crucial social systems---ranging from crime fighting to job portals to hospital---it is recommended that social-systems analysis should also include the possible effects of AI on their performance throughout the various stages of the design cycle (i.e. conception, design, implementation, and deployment) in social institutions.


BD4D practitioners should follow the lead of the following five principles of data for humanity laid down in \cite{d4humanity}: \textit{(1) Do no harm; (2) Use data to help create peaceful coexistence; (3) Use data to help vulnerable people and people in need; (4) Use data to preserve and improve the natural environment; and (5) Use data to help create a fair world without discrimination}. 

\vspace{2mm}
BD4D practitioners will also do well to adhere to the following oaths, which were developed by Herman and Wilmott as the ``Modelers' Hippocratic Oath'' \cite{derman2009financial} in the light of the global financial crisis: \textit{(1) Though I will use models boldly to estimate value, I will not be overly impressed by mathematics; (2) I will never sacrifice reality for elegance without explaining why I have done so; (3) Nor will I give the people who use my model false comfort about its accuracy. Instead, I will make explicit its assumptions and oversights; and (4) I understand that my work may have enormous effects on society and the economy, many of them beyond my comprehension.}



\section{Conclusions}

In this paper, we have provided a cautious perspective on the use of big data for human development. While we believe that big data has great potential for facilitating human development, our aim is to caution against an uncritical acceptance and careless application of big data methods in matters directly affecting human welfare and development. We need to guard against a na\"ive overreliance on data to avoid the many pitfalls of data worship. We argue that big data technology is a tool, and like all tools it should be considered as a handmaiden rather than as a headmaster. In particular, we argue that big data analytics cannot substitute for good research design and subject-matter knowledge. Various potential remedies to address the various pitfalls of using big data for development have also been highlighted. To conclude, we will like to emphasize that our paper should certainly not be construed as a techno-phobic manifesto; we believe strongly in the promise of big data for development (BD4D) but when pursued with due attention to the mitigation of the many associated pitfalls. 


\bibliographystyle{plain}
\bibliography{refs}

\end{document}